\documentclass[conference]{IEEEtran}

\IEEEoverridecommandlockouts
\usepackage{booktabs}
 \usepackage[table,xcdraw]{xcolor}
\usepackage[hidelinks]{hyperref}
\usepackage{wrapfig}
\usepackage{verbatim} 
\usepackage{amsmath,amssymb,amsthm} 
\usepackage{listings} 
\usepackage{color}
\usepackage{graphicx}
\usepackage{enumerate}
\usepackage{paralist}

\usepackage[oldcommands,linesnumbered,boxed]{algorithm2e}
\setalcapskip{1em} 
\newcommand{\martinAlgFontsize}{\fontsize{8.0}{8.0}\selectfont}
\SetAlFnt{\martinAlgFontsize}
\newcommand{\martinAlgCapFontsize}{\fontsize{9.0}{9.0}\selectfont}
\SetAlCapNameFnt{\martinAlgCapFontsize}
\SetAlCapFnt{\martinAlgCapFontsize}

\newcommand{\MSCS}{MPMCS}
\newcommand{\tool}{MPMCS4FTA}

\newcommand{\ackContent}{
    This work has been supported by the European Union's Horizon 2020 research and innovation programme under grant No 739551 (KIOS CoE). 
    To appear in Proceedings of the MaxSAT Evaluation 2020 (MSE'20),  \hbox{\url{https://maxsat-evaluations.github.io/2020/}}.
}

\begin{document}

\thispagestyle{plain}
\pagestyle{plain}

\title{MaxSAT Evaluation 2020 - Benchmark: \\Identifying Maximum Probability Minimal Cut Sets in Fault Trees
		\vspace{-0.2cm}}

\author{   
	\thanks{\ackContent}
	\IEEEauthorblockN{
		Mart\'in Barr\`ere and         
		Chris Hankin
    }
	\IEEEauthorblockA{Institute for Security Science and Technology, Imperial College London, UK
		\\\{m.barrere, c.hankin\}@imperial.ac.uk}	
		\vspace{-0.7cm}
}

\maketitle

\begin{abstract}
This paper presents a MaxSAT benchmark focused on the identification of Maximum Probability Minimal Cut Sets (\MSCS s) in fault trees. 
We address the \MSCS~problem by transforming the input fault tree into a weighted logical formula that is then used to build and solve a Weighted Partial MaxSAT problem. The benchmark includes 80 cases with fault trees of different size and composition as well as the optimal cost and solution for each case. 
\end{abstract}


\newcommand{\content}{sections} 

\section{Problem overview}
\label{sec:intro}

Fault Tree Analysis (FTA) is an analytical tool aimed at modelling and evaluating how complex systems may fail. FTA is widely used as a risk assessment tool in safety and reliability engineering for a broad range of industries including aerospace, power plants, nuclear plants, and others high-hazard fields \cite{Ruijters2015}. Essentially, a fault tree is a directed acyclic graph~(DAG) which involves a set of basic events (e.g. component failures) that are combined using logic operators (e.g. AND and OR gates) to model how these events may lead to an undesired state of the system normally represented at the root of the tree (top level event). 

Our work is focused on a novel measure for FTA in the form of a hybrid analysis technique that involves quantitative and qualitative aspects of fault trees. From a qualitative perspective, we focus on Minimal Cut Sets (MCS). An MCS is a minimal combination of events that together cause the top level event. As such, MCSs are fundamental for structural analysis. The problem is that, in large scenarios, computing all MCSs might be very expensive and there might be hundreds of MCSs, which makes it hard to handle and prioritise which MCSs should be attended first. 
In that context, the goal of this work is to identify the MCS with maximum probability. We call this problem the \MSCS. 
This is an NP-complete problem and we use a MaxSAT-based approach to address it.

\section{Simple example}

\begin{figure}[!h]
	\centering
	\includegraphics[scale=0.19]{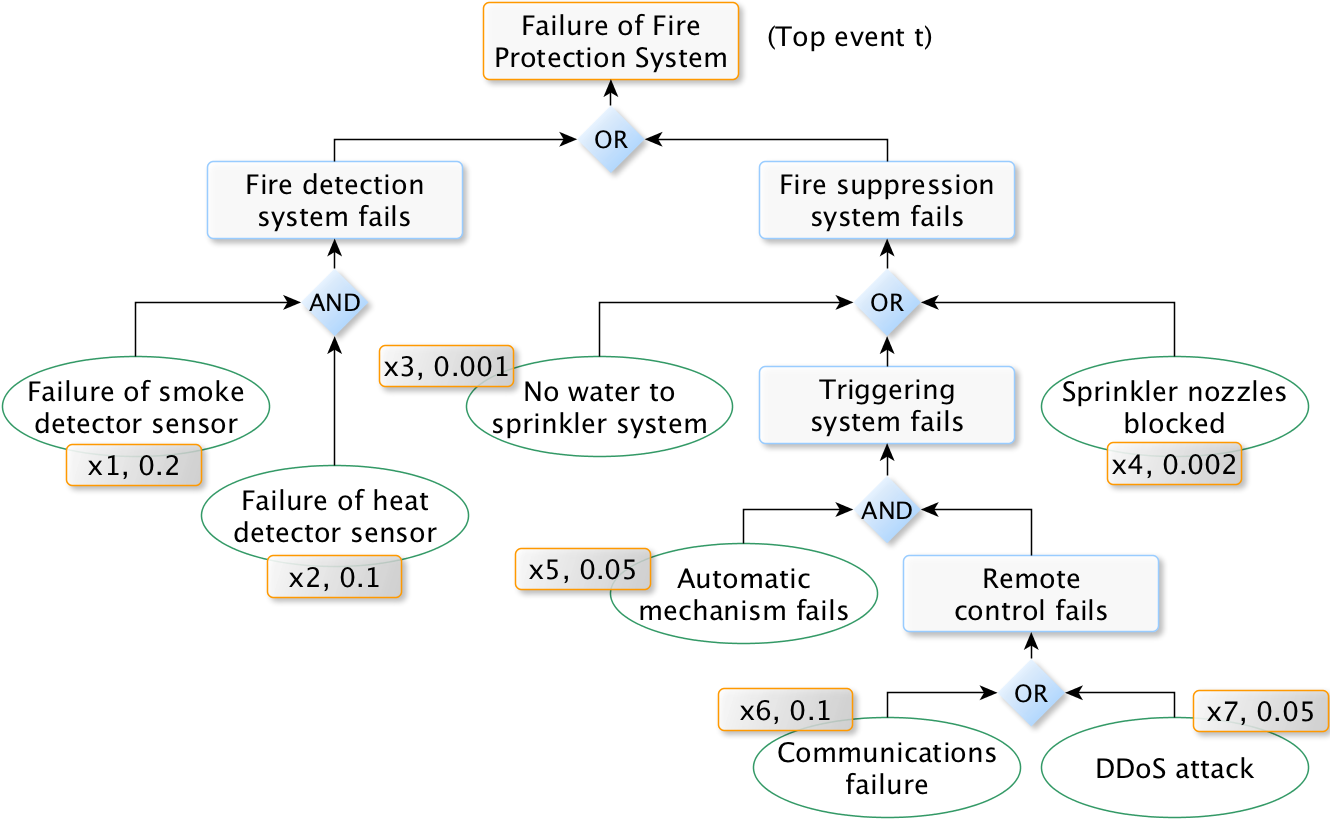}	
        \vspace{-0.6cm}
	\caption{Fault tree of a cyber-physical fire protection system (simplified)}
	\label{fig:simple-example1}
    \vspace{-0.15cm}
\end{figure}

The fault tree shown in Fig. \ref{fig:simple-example1} illustrates the different combinations of events that may lead to the failure of an hypothetical Fire Protection System (FPS) based on \cite{Kabir2017}. The FPS can fail if either the fire detection system or the fire suppression mechanism fails. In turn, the detection system can fail if both sensors fail simultaneously (events $x_1$ and $x_2$), while the suppression mechanism may fail if there is no water ($x3$), the sprinkler nozzles are blocked ($x_4$), or the triggering system does not work. The latter can fail if neither of its operation modes (automatic ($x_5$) or remotely operated) works properly. The remote control can fail if the communications channel fails ($x_6$) or the channel is not available due to a cyber attack, e.g. DDoS attack ($x_7$). Each basic event has an associated value that indicates its probability of occurrence~$p(x_i)$.

\vspace{-0.2cm}

A fault tree $F$ can be represented as a Boolean equation $f(t)$ that expresses the different ways in which the top event $t$ can be satisfied \cite{FtaHandbook2002}. 
In our example, $f(t)$ is as follows: 

\vspace{-0.5cm}
	\begin{equation}
	\begin{array}{c}
	f(t) = (x_1 \land x_2) \lor (x_3 \lor x_4 \lor (x_5 \land (x_6 \lor x_7))) \nonumber
	\end{array} 
	\vspace{-0.1cm}
	\end{equation}

The objective is to find the minimal set of logical variables that makes the equation $f(t)$ \textit{true} and whose joint probability is maximal among all minimal sets. In our example, the \MSCS~is $\{x_1,x_2\}$ with a joint probability of $0.02$. 

\vspace{0.15cm}
\section{MaxSAT formulation strategy}

Given a fault tree and its logical formulation $f(t)$, we carry out a series of steps to compute the \MSCS~as follows. 

\textbf{1. Logical transformation.} Since we are interested in minimising the number of satisfied clauses, which is opposed to what MaxSAT does (maximisation), we flip all logic gates but keep all events in their positive form. 
In our example, we obtain: $g(t) = (x_1 \lor x_2) \land (x_3 \land x_4 \land (x_5 \lor (x_6 \land x_7)))$. 
Then, the objective is to satisfy $\neg g(t)$ where the falsified variables will indicate the minimum set of events that must simultaneously occur to trigger the top level event. 
A more detailed explanation of this transformation can be found in \cite{Barrere-Arxiv2020-FTA}. 
We then use the Tseitin transformation to produce in polynomial time an equisatisfiable CNF formula \cite{Tseitin70}. 
    
\textbf{2. MaxSAT weights.} 
Due to the fact that MaxSAT is additive in nature and the \MSCS~problem involves the multiplication of decision variables, we transform the probabilities into a negative log-space so the multiplication becomes a sum. In addition, many SAT solvers only support integer weights so we perform a second transformation by right shifting (multiplying by 10) every value until the smallest value is covered with an acceptable level of precision. For example, 0.001 and 0.00007 would become 100 and 7 (right shift 5 times). 
Additional variables introduced by the Tseitin transformation have weight 0. 
We then specify the problem as a Partial Weighted MaxSAT instance  by assigning the transformed probability values as a penalty score for each decision variable. 
    
\textbf{3. Parallel SAT-solving architecture.} 
Since different SAT solvers normally use different resolution techniques, some of them are very good at some instances and not that good at others. To address this issue, we run multiple SAT-solvers in parallel and pick the solution of the solver that finishes first. 
We have experimentally observed that the combination of different solvers provides good results in terms of performance and scalability. 
Once the solution has been found, we translate back the transformed values into their stochastic domain and output the MCS with maximum probability.

\section{Fault tree generation} 
The benchmark presented in this paper relies on our open source tool \tool~\cite{BarrereMPMCS4FTAGithub}. 
We have used \tool~to generate and analyse synthetic pseudo-random fault trees of different size and composition. 
We use AND/OR graphs as the underlying structure to represent fault trees. 
The benchmark presented in \cite{Barrere-MaxSAT-Benchmark-Arxiv2019} also considers AND/OR graphs as a means to represent operational dependencies between components in industrial control systems \cite{BarrereJisa2020}.  
However, the instances presented in this paper differ in that: 
\begin{inparaenum}
    \item they are restricted to directed acyclic graphs (DAGs), 
    \item only the basic events represented at the leaves of the fault tree involve a probability of failure, and
    \item leaves can have more than one parent in order to relax the definition of strict logical trees. 
\end{inparaenum}

We control the size and composition of a random fault tree of size $n$ according to a configuration $R=(R_{AT},R_{AND},R_{OR})$. 
$R_{AT} \in [0,1]$ indicates the proportion of atomic nodes (basic events) with respect to size $n$ (e.g. $0.2$ means $20\%$) whereas 
$R_{AND}$ and $R_{OR}$ indicate the proportion of AND and OR nodes respectively. 
To create a fault tree of size $n$, we first create two lists: $L=\{l_1, \ldots, l_m\}$ and $A=\{a_1,\ldots,a_s\}$. 
$L$ is a random sequence of AND and OR nodes with the specified proportions for each operator where $m=n*(R_{AND}+R_{OR})$. 
$A$ is a list of atomic nodes where $s=n*R_{AT}$, thus $n=m+s$. 
In addition, each atomic node has a random probability of failure $p(a_i) \in [0,1]$. 

To ensure connectivity, we first create the root node $t$ and connect $l_1$ to $t$ ($l_1 \rightarrow t$). 
Then, for each logic node $l_i$ in the sequence $L$, we randomly choose $k$ nodes $l_j$ ahead (thus $j>i$) and create $k$ edges ($l_j \rightarrow l_i$) in the tree. 
When the remaining nodes in $L$ are not enough to cover $k$ nodes, we use random atomic nodes from $A$. 
At this point, we also make sure that $l_i$ points to at least one previous node in the sequence $L$. If that is not the case, we choose a random node $l_h$ (with $h < i$) and create an edge ($l_i \rightarrow l_h$). 
Once the sequence $L$ has been processed, we traverse the list $A$ and connect each atomic node $a_i$ as follows. 
First, we draw a random value $k'$ between 1 and $k$. Then, we add random edges ($a_i \rightarrow l_j$) from $a_i$ to logic nodes $l_j$  until we cover $k'$ connections.

\vspace{-0.15cm}
\section{Benchmark description}

Out dataset includes 80 cases in total, and can be obtained at \cite{BarrereMPMCS4FTAGithub}. 
It contains fault trees with four different sizes: 2500, 5000, 7500, and 10000 nodes (20 cases each). 
For each tree size, we consider two different graph configurations, $R_1=(0.8,0.1,0.1)$ and $R_2=(0.6,0.2,0.2)$, which determine the composition of the fault trees (10 cases each).  
Table \ref{tab:experiments} shows the identifiers of the cases within each one of these categories. 

\vspace{-0.1cm}
\begin{table}[h!]
    \centering
    \begin{tabular}{@{}|c|c|c|@{}}
        \toprule
        \textbf{\#Nodes/Configurations} & \textbf{$R_1=(0.8,0.1,0.1)$} & \textbf{$R_2=(0.6,0.2,0.2)$} \\ \midrule
        2500 & 1 to 10 & 11 to 20 \\ \midrule
        5000 & 21 to 30 & 31 to 40 \\ \midrule
        7500 & 41 to 50 & 51 to 60 \\ \midrule
        10000 & 61 to 70 & 71 to 80 \\ \bottomrule
    \end{tabular}
    \vspace{0.05cm}
    \caption{Benchmark cases and configurations}
    \label{tab:experiments}
\end{table}
\vspace{-0.5cm}
Each case is specified in an individual \textbf{.wcnf} (DIMACS-like, weighted CNF) file named with the case id and the number of nodes involved. 
The weight for hard clauses (\textit{top} value) has been set to $2.0 \times 10^9$. 
The weight of each soft constraint is an integer value that corresponds to the transformation (right shifting) of the probability value in $-log$ space. 
Tables \ref{tab:part1} and \ref{tab:part2} detail each case as well as the results obtained with our tool. 
The field \textbf{id} identifies each case.  
\textbf{gNodes} and \textbf{gEdges} indicate the total number of nodes and edges in the fault tree. 
\textbf{gAT}, \textbf{gAND}, and \textbf{gOR}, indicate the approximate composition of the graph in terms of atomic (basic events), AND, and OR nodes.  
\textbf{tsVars} and \textbf{tsClauses} show the number of variables and clauses involved in the MaxSAT formulation after applying the Tseitin transformation.   
\textbf{time} shows the resolution time reported by \tool~in milliseconds. 
Currently, the MaxSAT solvers used in \tool~are SAT4J~\cite{SAT4J} and a Python-based linear programming approach using Gurobi~\cite{Gurobi}. 
\textbf{size} indicates the number of nodes identified in the \MSCS~solution. 
\textbf{intLogCost} indicates the cost of the solution in $-log$ space as an integer value (right shifted). 
\textbf{logCost} indicates the cost of the solution in $-log$ space. 
\textbf{\MSCS~probability} indicates the joint probability of the \MSCS. 
These experiments have been performed on a MacBook Pro (16-inch, 2019), 2.4 GHz 8-core Intel Core i9, 32 GB 2666 MHz DDR4.

\begin{table*}[]
	\centering
\begin{tabular}{@{}ccccccccccccc@{}}
\toprule
\multicolumn{1}{|c|}{\textbf{id}} & \multicolumn{1}{c|}{\textbf{gNodes}} & \multicolumn{1}{c|}{\textbf{gEdges}} & \multicolumn{1}{c|}{\textbf{gAT}} & \multicolumn{1}{c|}{\textbf{gAND}} & \multicolumn{1}{c|}{\textbf{gOR}} & \multicolumn{1}{c|}{\textbf{tsVars}} & \multicolumn{1}{c|}{\textbf{tsClauses}} & \multicolumn{1}{c|}{\textbf{time}} & \multicolumn{1}{c|}{\textbf{size}} & \multicolumn{1}{c|}{\textbf{intLogCost}} & \multicolumn{1}{c|}{\textbf{logCost}} & \multicolumn{1}{c|}{\textbf{MPMCS probability}}    \\ \midrule
\multicolumn{1}{|c|}{1}           & \multicolumn{1}{c|}{2500}            & \multicolumn{1}{c|}{7151}            & \multicolumn{1}{c|}{2002}         & \multicolumn{1}{c|}{250}           & \multicolumn{1}{c|}{250}          & \multicolumn{1}{c|}{1978}            & \multicolumn{1}{c|}{6258}               & \multicolumn{1}{c|}{618}           & \multicolumn{1}{c|}{1}                    & \multicolumn{1}{c|}{246}                        & \multicolumn{1}{c|}{2.46E-4}                 & \multicolumn{1}{c|}{0.999754}                \\ \midrule
\multicolumn{1}{|c|}{2}           & \multicolumn{1}{c|}{2500}            & \multicolumn{1}{c|}{7192}            & \multicolumn{1}{c|}{2002}         & \multicolumn{1}{c|}{250}           & \multicolumn{1}{c|}{250}          & \multicolumn{1}{c|}{4268}            & \multicolumn{1}{c|}{15026}              & \multicolumn{1}{c|}{850}           & \multicolumn{1}{c|}{447}                  & \multicolumn{1}{c|}{464771733}                  & \multicolumn{1}{c|}{464.771733}              & \multicolumn{1}{c|}{1.42239870668983E-202}   \\ \midrule
\multicolumn{1}{|c|}{3}           & \multicolumn{1}{c|}{2500}            & \multicolumn{1}{c|}{7196}            & \multicolumn{1}{c|}{2002}         & \multicolumn{1}{c|}{250}           & \multicolumn{1}{c|}{250}          & \multicolumn{1}{c|}{1207}            & \multicolumn{1}{c|}{3763}               & \multicolumn{1}{c|}{290}           & \multicolumn{1}{c|}{1}                    & \multicolumn{1}{c|}{27591}                      & \multicolumn{1}{c|}{0.027591}                & \multicolumn{1}{c|}{0.972787}                \\ \midrule
\multicolumn{1}{|c|}{4}           & \multicolumn{1}{c|}{2500}            & \multicolumn{1}{c|}{7140}            & \multicolumn{1}{c|}{2002}         & \multicolumn{1}{c|}{250}           & \multicolumn{1}{c|}{250}          & \multicolumn{1}{c|}{4211}            & \multicolumn{1}{c|}{14673}              & \multicolumn{1}{c|}{833}           & \multicolumn{1}{c|}{1}                    & \multicolumn{1}{c|}{238}                        & \multicolumn{1}{c|}{2.38E-4}                 & \multicolumn{1}{c|}{0.999763}                \\ \midrule
\multicolumn{1}{|c|}{5}           & \multicolumn{1}{c|}{2500}            & \multicolumn{1}{c|}{7107}            & \multicolumn{1}{c|}{2002}         & \multicolumn{1}{c|}{250}           & \multicolumn{1}{c|}{250}          & \multicolumn{1}{c|}{3907}            & \multicolumn{1}{c|}{13325}              & \multicolumn{1}{c|}{821}           & \multicolumn{1}{c|}{1}                    & \multicolumn{1}{c|}{7879}                       & \multicolumn{1}{c|}{0.007879}                & \multicolumn{1}{c|}{0.992153}                \\ \midrule
\multicolumn{1}{|c|}{6}           & \multicolumn{1}{c|}{2500}            & \multicolumn{1}{c|}{7202}            & \multicolumn{1}{c|}{2002}         & \multicolumn{1}{c|}{250}           & \multicolumn{1}{c|}{250}          & \multicolumn{1}{c|}{3410}            & \multicolumn{1}{c|}{11350}              & \multicolumn{1}{c|}{749}           & \multicolumn{1}{c|}{70}                   & \multicolumn{1}{c|}{81474531}                   & \multicolumn{1}{c|}{81.474531}               & \multicolumn{1}{c|}{4.147681160335815E-36}   \\ \midrule
\multicolumn{1}{|c|}{7}           & \multicolumn{1}{c|}{2500}            & \multicolumn{1}{c|}{7126}            & \multicolumn{1}{c|}{2002}         & \multicolumn{1}{c|}{250}           & \multicolumn{1}{c|}{250}          & \multicolumn{1}{c|}{3304}            & \multicolumn{1}{c|}{10922}              & \multicolumn{1}{c|}{711}           & \multicolumn{1}{c|}{1}                    & \multicolumn{1}{c|}{315}                        & \multicolumn{1}{c|}{3.15E-4}                 & \multicolumn{1}{c|}{0.999685}                \\ \midrule
\multicolumn{1}{|c|}{8}           & \multicolumn{1}{c|}{2500}            & \multicolumn{1}{c|}{7181}            & \multicolumn{1}{c|}{2002}         & \multicolumn{1}{c|}{250}           & \multicolumn{1}{c|}{250}          & \multicolumn{1}{c|}{3752}            & \multicolumn{1}{c|}{12713}              & \multicolumn{1}{c|}{826}           & \multicolumn{1}{c|}{1}                    & \multicolumn{1}{c|}{2576}                       & \multicolumn{1}{c|}{0.002576}                & \multicolumn{1}{c|}{0.997428}                \\ \midrule
\multicolumn{1}{|c|}{9}           & \multicolumn{1}{c|}{2500}            & \multicolumn{1}{c|}{7157}            & \multicolumn{1}{c|}{2002}         & \multicolumn{1}{c|}{250}           & \multicolumn{1}{c|}{250}          & \multicolumn{1}{c|}{3011}            & \multicolumn{1}{c|}{9847}               & \multicolumn{1}{c|}{625}           & \multicolumn{1}{c|}{1}                    & \multicolumn{1}{c|}{4301}                       & \multicolumn{1}{c|}{0.004301}                & \multicolumn{1}{c|}{0.995709}                \\ \midrule
\multicolumn{1}{|c|}{10}          & \multicolumn{1}{c|}{2500}            & \multicolumn{1}{c|}{7156}            & \multicolumn{1}{c|}{2002}         & \multicolumn{1}{c|}{250}           & \multicolumn{1}{c|}{250}          & \multicolumn{1}{c|}{642}             & \multicolumn{1}{c|}{1982}               & \multicolumn{1}{c|}{211}           & \multicolumn{1}{c|}{19}                   & \multicolumn{1}{c|}{12423488}                   & \multicolumn{1}{c|}{12.423488}               & \multicolumn{1}{c|}{4.0231156723921624E-6}   \\ \midrule
\multicolumn{1}{|c|}{11}          & \multicolumn{1}{c|}{2500}            & \multicolumn{1}{c|}{6831}            & \multicolumn{1}{c|}{1502}         & \multicolumn{1}{c|}{500}           & \multicolumn{1}{c|}{500}          & \multicolumn{1}{c|}{3873}            & \multicolumn{1}{c|}{14170}              & \multicolumn{1}{c|}{912}           & \multicolumn{1}{c|}{1}                    & \multicolumn{1}{c|}{28842}                      & \multicolumn{1}{c|}{0.028842}                & \multicolumn{1}{c|}{0.971571}                \\ \midrule
\multicolumn{1}{|c|}{12}          & \multicolumn{1}{c|}{2500}            & \multicolumn{1}{c|}{6782}            & \multicolumn{1}{c|}{1502}         & \multicolumn{1}{c|}{500}           & \multicolumn{1}{c|}{500}          & \multicolumn{1}{c|}{2377}            & \multicolumn{1}{c|}{7941}               & \multicolumn{1}{c|}{550}           & \multicolumn{1}{c|}{1}                    & \multicolumn{1}{c|}{32680}                      & \multicolumn{1}{c|}{0.03268}                 & \multicolumn{1}{c|}{0.96785}                 \\ \midrule
\multicolumn{1}{|c|}{13}          & \multicolumn{1}{c|}{2500}            & \multicolumn{1}{c|}{6814}            & \multicolumn{1}{c|}{1502}         & \multicolumn{1}{c|}{500}           & \multicolumn{1}{c|}{500}          & \multicolumn{1}{c|}{3216}            & \multicolumn{1}{c|}{11235}              & \multicolumn{1}{c|}{700}           & \multicolumn{1}{c|}{13}                   & \multicolumn{1}{c|}{10769787}                   & \multicolumn{1}{c|}{10.769787}               & \multicolumn{1}{c|}{2.1025796252653052E-5}   \\ \midrule
\multicolumn{1}{|c|}{14}          & \multicolumn{1}{c|}{2500}            & \multicolumn{1}{c|}{6700}            & \multicolumn{1}{c|}{1502}         & \multicolumn{1}{c|}{500}           & \multicolumn{1}{c|}{500}          & \multicolumn{1}{c|}{3268}            & \multicolumn{1}{c|}{11376}              & \multicolumn{1}{c|}{728}           & \multicolumn{1}{c|}{197}                  & \multicolumn{1}{c|}{207945092}                  & \multicolumn{1}{c|}{207.945092}              & \multicolumn{1}{c|}{4.9088521396478804E-91}  \\ \midrule
\multicolumn{1}{|c|}{15}          & \multicolumn{1}{c|}{2500}            & \multicolumn{1}{c|}{6897}            & \multicolumn{1}{c|}{1502}         & \multicolumn{1}{c|}{500}           & \multicolumn{1}{c|}{500}          & \multicolumn{1}{c|}{3063}            & \multicolumn{1}{c|}{10555}              & \multicolumn{1}{c|}{817}           & \multicolumn{1}{c|}{1}                    & \multicolumn{1}{c|}{3262}                       & \multicolumn{1}{c|}{0.003262}                & \multicolumn{1}{c|}{0.996744}                \\ \midrule
\multicolumn{1}{|c|}{16}          & \multicolumn{1}{c|}{2500}            & \multicolumn{1}{c|}{6849}            & \multicolumn{1}{c|}{1502}         & \multicolumn{1}{c|}{500}           & \multicolumn{1}{c|}{500}          & \multicolumn{1}{c|}{2044}            & \multicolumn{1}{c|}{6765}               & \multicolumn{1}{c|}{470}           & \multicolumn{1}{c|}{1}                    & \multicolumn{1}{c|}{191116}                     & \multicolumn{1}{c|}{0.191116}                & \multicolumn{1}{c|}{0.826037}                \\ \midrule
\multicolumn{1}{|c|}{17}          & \multicolumn{1}{c|}{2500}            & \multicolumn{1}{c|}{6787}            & \multicolumn{1}{c|}{1502}         & \multicolumn{1}{c|}{500}           & \multicolumn{1}{c|}{500}          & \multicolumn{1}{c|}{3158}            & \multicolumn{1}{c|}{10955}              & \multicolumn{1}{c|}{723}           & \multicolumn{1}{c|}{1}                    & \multicolumn{1}{c|}{284520}                     & \multicolumn{1}{c|}{0.28452}                 & \multicolumn{1}{c|}{0.752376}                \\ \midrule
\multicolumn{1}{|c|}{18}          & \multicolumn{1}{c|}{2500}            & \multicolumn{1}{c|}{6872}            & \multicolumn{1}{c|}{1502}         & \multicolumn{1}{c|}{500}           & \multicolumn{1}{c|}{500}          & \multicolumn{1}{c|}{3433}            & \multicolumn{1}{c|}{12147}              & \multicolumn{1}{c|}{773}           & \multicolumn{1}{c|}{139}                  & \multicolumn{1}{c|}{130484455}                  & \multicolumn{1}{c|}{130.484455}              & \multicolumn{1}{c|}{2.1453798325228181E-57}  \\ \midrule
\multicolumn{1}{|c|}{19}          & \multicolumn{1}{c|}{2500}            & \multicolumn{1}{c|}{6821}            & \multicolumn{1}{c|}{1502}         & \multicolumn{1}{c|}{500}           & \multicolumn{1}{c|}{500}          & \multicolumn{1}{c|}{2506}            & \multicolumn{1}{c|}{8439}               & \multicolumn{1}{c|}{534}           & \multicolumn{1}{c|}{17}                   & \multicolumn{1}{c|}{9662887}                    & \multicolumn{1}{c|}{9.662887}                & \multicolumn{1}{c|}{6.36019885647539E-5}     \\ \midrule
\multicolumn{1}{|c|}{20}          & \multicolumn{1}{c|}{2500}            & \multicolumn{1}{c|}{6831}            & \multicolumn{1}{c|}{1502}         & \multicolumn{1}{c|}{500}           & \multicolumn{1}{c|}{500}          & \multicolumn{1}{c|}{3848}            & \multicolumn{1}{c|}{14095}              & \multicolumn{1}{c|}{821}           & \multicolumn{1}{c|}{1}                    & \multicolumn{1}{c|}{3507}                       & \multicolumn{1}{c|}{0.003507}                & \multicolumn{1}{c|}{0.996501}                \\ \midrule
\multicolumn{1}{|c|}{21}          & \multicolumn{1}{c|}{5000}            & \multicolumn{1}{c|}{14324}           & \multicolumn{1}{c|}{4002}         & \multicolumn{1}{c|}{500}           & \multicolumn{1}{c|}{500}          & \multicolumn{1}{c|}{4149}            & \multicolumn{1}{c|}{13224}              & \multicolumn{1}{c|}{932}           & \multicolumn{1}{c|}{229}                  & \multicolumn{1}{c|}{217397271}                  & \multicolumn{1}{c|}{217.397271}              & \multicolumn{1}{c|}{3.8565352927569054E-95}  \\ \midrule
\multicolumn{1}{|c|}{22}          & \multicolumn{1}{c|}{5000}            & \multicolumn{1}{c|}{14313}           & \multicolumn{1}{c|}{4002}         & \multicolumn{1}{c|}{500}           & \multicolumn{1}{c|}{500}          & \multicolumn{1}{c|}{8532}            & \multicolumn{1}{c|}{29961}              & \multicolumn{1}{c|}{925}           & \multicolumn{1}{c|}{614}                  & \multicolumn{1}{c|}{641968767}                  & \multicolumn{1}{c|}{641.968767}              & \multicolumn{1}{c|}{1.5912873405576694E-279} \\ \midrule
\multicolumn{1}{|c|}{23}          & \multicolumn{1}{c|}{5000}            & \multicolumn{1}{c|}{14329}           & \multicolumn{1}{c|}{4002}         & \multicolumn{1}{c|}{500}           & \multicolumn{1}{c|}{500}          & \multicolumn{1}{c|}{6971}            & \multicolumn{1}{c|}{23338}              & \multicolumn{1}{c|}{842}           & \multicolumn{1}{c|}{240}                  & \multicolumn{1}{c|}{251915559}                  & \multicolumn{1}{c|}{251.915559}              & \multicolumn{1}{c|}{3.9351584673463555E-110} \\ \midrule
\multicolumn{1}{|c|}{24}          & \multicolumn{1}{c|}{5000}            & \multicolumn{1}{c|}{14361}           & \multicolumn{1}{c|}{4002}         & \multicolumn{1}{c|}{500}           & \multicolumn{1}{c|}{500}          & \multicolumn{1}{c|}{8020}            & \multicolumn{1}{c|}{27645}              & \multicolumn{1}{c|}{843}           & \multicolumn{1}{c|}{1}                    & \multicolumn{1}{c|}{793}                        & \multicolumn{1}{c|}{7.93E-4}                 & \multicolumn{1}{c|}{0.999209}                \\ \midrule
\multicolumn{1}{|c|}{25}          & \multicolumn{1}{c|}{5000}            & \multicolumn{1}{c|}{14370}           & \multicolumn{1}{c|}{4002}         & \multicolumn{1}{c|}{500}           & \multicolumn{1}{c|}{500}          & \multicolumn{1}{c|}{8965}            & \multicolumn{1}{c|}{32190}              & \multicolumn{1}{c|}{843}           & \multicolumn{1}{c|}{1}                    & \multicolumn{1}{c|}{1858}                       & \multicolumn{1}{c|}{0.001858}                & \multicolumn{1}{c|}{0.998144}                \\ \midrule
\multicolumn{1}{|c|}{26}          & \multicolumn{1}{c|}{5000}            & \multicolumn{1}{c|}{14317}           & \multicolumn{1}{c|}{4002}         & \multicolumn{1}{c|}{500}           & \multicolumn{1}{c|}{500}          & \multicolumn{1}{c|}{5443}            & \multicolumn{1}{c|}{17581}              & \multicolumn{1}{c|}{827}           & \multicolumn{1}{c|}{1}                    & \multicolumn{1}{c|}{3615}                       & \multicolumn{1}{c|}{0.003615}                & \multicolumn{1}{c|}{0.996391}                \\ \midrule
\multicolumn{1}{|c|}{27}          & \multicolumn{1}{c|}{5000}            & \multicolumn{1}{c|}{14407}           & \multicolumn{1}{c|}{4002}         & \multicolumn{1}{c|}{500}           & \multicolumn{1}{c|}{500}          & \multicolumn{1}{c|}{8113}            & \multicolumn{1}{c|}{28023}              & \multicolumn{1}{c|}{842}           & \multicolumn{1}{c|}{277}                  & \multicolumn{1}{c|}{253971185}                  & \multicolumn{1}{c|}{253.971185}              & \multicolumn{1}{c|}{5.035082961027143E-111}  \\ \midrule
\multicolumn{1}{|c|}{28}          & \multicolumn{1}{c|}{5000}            & \multicolumn{1}{c|}{14365}           & \multicolumn{1}{c|}{4002}         & \multicolumn{1}{c|}{500}           & \multicolumn{1}{c|}{500}          & \multicolumn{1}{c|}{8952}            & \multicolumn{1}{c|}{32153}              & \multicolumn{1}{c|}{837}           & \multicolumn{1}{c|}{1041}                 & \multicolumn{1}{c|}{994658460}                  & \multicolumn{1}{c|}{994.65846}               & \multicolumn{1}{c|}{0.0}                     \\ \midrule
\multicolumn{1}{|c|}{29}          & \multicolumn{1}{c|}{5000}            & \multicolumn{1}{c|}{14321}           & \multicolumn{1}{c|}{4002}         & \multicolumn{1}{c|}{500}           & \multicolumn{1}{c|}{500}          & \multicolumn{1}{c|}{8859}            & \multicolumn{1}{c|}{31477}              & \multicolumn{1}{c|}{833}           & \multicolumn{1}{c|}{379}                  & \multicolumn{1}{c|}{378308687}                  & \multicolumn{1}{c|}{378.308687}              & \multicolumn{1}{c|}{5.051735441001231E-165}  \\ \midrule
\multicolumn{1}{|c|}{30}          & \multicolumn{1}{c|}{5000}            & \multicolumn{1}{c|}{14316}           & \multicolumn{1}{c|}{4002}         & \multicolumn{1}{c|}{500}           & \multicolumn{1}{c|}{500}          & \multicolumn{1}{c|}{7948}            & \multicolumn{1}{c|}{27315}              & \multicolumn{1}{c|}{830}           & \multicolumn{1}{c|}{1}                    & \multicolumn{1}{c|}{970}                        & \multicolumn{1}{c|}{9.7E-4}                  & \multicolumn{1}{c|}{0.999032}                \\ \midrule
\multicolumn{1}{|c|}{31}          & \multicolumn{1}{c|}{5000}            & \multicolumn{1}{c|}{13607}           & \multicolumn{1}{c|}{3002}         & \multicolumn{1}{c|}{1000}          & \multicolumn{1}{c|}{1000}         & \multicolumn{1}{c|}{6384}            & \multicolumn{1}{c|}{22218}              & \multicolumn{1}{c|}{938}           & \multicolumn{1}{c|}{1}                    & \multicolumn{1}{c|}{2530}                       & \multicolumn{1}{c|}{0.00253}                 & \multicolumn{1}{c|}{0.997474}                \\ \midrule
\multicolumn{1}{|c|}{32}          & \multicolumn{1}{c|}{5000}            & \multicolumn{1}{c|}{13730}           & \multicolumn{1}{c|}{3002}         & \multicolumn{1}{c|}{1000}          & \multicolumn{1}{c|}{1000}         & \multicolumn{1}{c|}{7330}            & \multicolumn{1}{c|}{26390}              & \multicolumn{1}{c|}{863}           & \multicolumn{1}{c|}{65}                   & \multicolumn{1}{c|}{63984958}                   & \multicolumn{1}{c|}{63.984958}               & \multicolumn{1}{c|}{1.62844121698006E-28}    \\ \midrule
\multicolumn{1}{|c|}{33}          & \multicolumn{1}{c|}{5000}            & \multicolumn{1}{c|}{13687}           & \multicolumn{1}{c|}{3002}         & \multicolumn{1}{c|}{1000}          & \multicolumn{1}{c|}{1000}         & \multicolumn{1}{c|}{3181}            & \multicolumn{1}{c|}{10354}              & \multicolumn{1}{c|}{683}           & \multicolumn{1}{c|}{1}                    & \multicolumn{1}{c|}{25289}                      & \multicolumn{1}{c|}{0.025289}                & \multicolumn{1}{c|}{0.975029}                \\ \midrule
\multicolumn{1}{|c|}{34}          & \multicolumn{1}{c|}{5000}            & \multicolumn{1}{c|}{13600}           & \multicolumn{1}{c|}{3002}         & \multicolumn{1}{c|}{1000}          & \multicolumn{1}{c|}{1000}         & \multicolumn{1}{c|}{6293}            & \multicolumn{1}{c|}{21870}              & \multicolumn{1}{c|}{834}           & \multicolumn{1}{c|}{407}                  & \multicolumn{1}{c|}{424495269}                  & \multicolumn{1}{c|}{424.495269}              & \multicolumn{1}{c|}{4.413071223454673E-185}  \\ \midrule
\multicolumn{1}{|c|}{35}          & \multicolumn{1}{c|}{5000}            & \multicolumn{1}{c|}{13712}           & \multicolumn{1}{c|}{3002}         & \multicolumn{1}{c|}{1000}          & \multicolumn{1}{c|}{1000}         & \multicolumn{1}{c|}{7361}            & \multicolumn{1}{c|}{26650}              & \multicolumn{1}{c|}{895}           & \multicolumn{1}{c|}{179}                  & \multicolumn{1}{c|}{171277203}                  & \multicolumn{1}{c|}{171.277203}              & \multicolumn{1}{c|}{4.1251154050451916E-75}  \\ \midrule
\multicolumn{1}{|c|}{36}          & \multicolumn{1}{c|}{5000}            & \multicolumn{1}{c|}{13709}           & \multicolumn{1}{c|}{3002}         & \multicolumn{1}{c|}{1000}          & \multicolumn{1}{c|}{1000}         & \multicolumn{1}{c|}{6231}            & \multicolumn{1}{c|}{21647}              & \multicolumn{1}{c|}{831}           & \multicolumn{1}{c|}{22}                   & \multicolumn{1}{c|}{19249301}                   & \multicolumn{1}{c|}{19.249301}               & \multicolumn{1}{c|}{4.366753474609794E-9}    \\ \midrule
\multicolumn{1}{|c|}{37}          & \multicolumn{1}{c|}{5000}            & \multicolumn{1}{c|}{13612}           & \multicolumn{1}{c|}{3002}         & \multicolumn{1}{c|}{1000}          & \multicolumn{1}{c|}{1000}         & \multicolumn{1}{c|}{6202}            & \multicolumn{1}{c|}{21523}              & \multicolumn{1}{c|}{931}           & \multicolumn{1}{c|}{257}                  & \multicolumn{1}{c|}{273826234}                  & \multicolumn{1}{c|}{273.826234}              & \multicolumn{1}{c|}{1.2035873310274229E-119} \\ \midrule
\multicolumn{1}{|c|}{38}          & \multicolumn{1}{c|}{5000}            & \multicolumn{1}{c|}{13664}           & \multicolumn{1}{c|}{3002}         & \multicolumn{1}{c|}{1000}          & \multicolumn{1}{c|}{1000}         & \multicolumn{1}{c|}{4482}            & \multicolumn{1}{c|}{14952}              & \multicolumn{1}{c|}{824}           & \multicolumn{1}{c|}{1}                    & \multicolumn{1}{c|}{4317}                       & \multicolumn{1}{c|}{0.004317}                & \multicolumn{1}{c|}{0.995693}                \\ \midrule
\multicolumn{1}{|c|}{39}          & \multicolumn{1}{c|}{5000}            & \multicolumn{1}{c|}{13631}           & \multicolumn{1}{c|}{3002}         & \multicolumn{1}{c|}{1000}          & \multicolumn{1}{c|}{1000}         & \multicolumn{1}{c|}{7395}            & \multicolumn{1}{c|}{26641}              & \multicolumn{1}{c|}{827}           & \multicolumn{1}{c|}{83}                   & \multicolumn{1}{c|}{89562456}                   & \multicolumn{1}{c|}{89.562456}               & \multicolumn{1}{c|}{1.2695246380697898E-39}  \\ \midrule
\multicolumn{1}{|c|}{40}          & \multicolumn{1}{c|}{5000}            & \multicolumn{1}{c|}{13641}           & \multicolumn{1}{c|}{3002}         & \multicolumn{1}{c|}{1000}          & \multicolumn{1}{c|}{1000}         & \multicolumn{1}{c|}{7825}            & \multicolumn{1}{c|}{28775}              & \multicolumn{1}{c|}{831}           & \multicolumn{1}{c|}{1}                    & \multicolumn{1}{c|}{5974}                       & \multicolumn{1}{c|}{0.005974}                & \multicolumn{1}{c|}{0.994045}                \\ 
 \bottomrule
\end{tabular}
 \vspace{0.5cm}
\caption{Benchmark description - Cases 1 to 40}
\label{tab:part1}
\end{table*}
\begin{table*}[]
	\centering
\begin{tabular}{@{}ccccccccccccc@{}}
\toprule
\multicolumn{1}{|c|}{\textbf{id}} & \multicolumn{1}{c|}{\textbf{gNodes}} & \multicolumn{1}{c|}{\textbf{gEdges}} & \multicolumn{1}{c|}{\textbf{gAT}} & \multicolumn{1}{c|}{\textbf{gAND}} & \multicolumn{1}{c|}{\textbf{gOR}} & \multicolumn{1}{c|}{\textbf{tsVars}} & \multicolumn{1}{c|}{\textbf{tsClauses}} & \multicolumn{1}{c|}{\textbf{time}} & \multicolumn{1}{c|}{\textbf{size}} & \multicolumn{1}{c|}{\textbf{intLogCost}} & \multicolumn{1}{c|}{\textbf{logCost}} & \multicolumn{1}{c|}{\textbf{MPMCS probability}}    \\ \midrule
\multicolumn{1}{|c|}{41}          & \multicolumn{1}{c|}{7500}            & \multicolumn{1}{c|}{21502}           & \multicolumn{1}{c|}{6002}         & \multicolumn{1}{c|}{750}           & \multicolumn{1}{c|}{750}          & \multicolumn{1}{c|}{8871}            & \multicolumn{1}{c|}{28951}              & \multicolumn{1}{c|}{965}           & \multicolumn{1}{c|}{1}                    & \multicolumn{1}{c|}{160}                        & \multicolumn{1}{c|}{1.6E-4}                  & \multicolumn{1}{c|}{0.999841}                \\ \midrule
\multicolumn{1}{|c|}{42}          & \multicolumn{1}{c|}{7500}            & \multicolumn{1}{c|}{21515}           & \multicolumn{1}{c|}{6002}         & \multicolumn{1}{c|}{750}           & \multicolumn{1}{c|}{750}          & \multicolumn{1}{c|}{7191}            & \multicolumn{1}{c|}{23069}              & \multicolumn{1}{c|}{852}           & \multicolumn{1}{c|}{1}                    & \multicolumn{1}{c|}{393}                        & \multicolumn{1}{c|}{3.93E-4}                 & \multicolumn{1}{c|}{0.999607}                \\ \midrule
\multicolumn{1}{|c|}{43}          & \multicolumn{1}{c|}{7500}            & \multicolumn{1}{c|}{21497}           & \multicolumn{1}{c|}{6002}         & \multicolumn{1}{c|}{750}           & \multicolumn{1}{c|}{750}          & \multicolumn{1}{c|}{5716}            & \multicolumn{1}{c|}{18114}              & \multicolumn{1}{c|}{843}           & \multicolumn{1}{c|}{1}                    & \multicolumn{1}{c|}{1095}                       & \multicolumn{1}{c|}{0.001095}                & \multicolumn{1}{c|}{0.998906}                \\ \midrule
\multicolumn{1}{|c|}{44}          & \multicolumn{1}{c|}{7500}            & \multicolumn{1}{c|}{21536}           & \multicolumn{1}{c|}{6002}         & \multicolumn{1}{c|}{750}           & \multicolumn{1}{c|}{750}          & \multicolumn{1}{c|}{6476}            & \multicolumn{1}{c|}{20645}              & \multicolumn{1}{c|}{849}           & \multicolumn{1}{c|}{600}                  & \multicolumn{1}{c|}{607247314}                  & \multicolumn{1}{c|}{607.247314}              & \multicolumn{1}{c|}{1.8912103369207186E-264} \\ \midrule
\multicolumn{1}{|c|}{45}          & \multicolumn{1}{c|}{7500}            & \multicolumn{1}{c|}{21472}           & \multicolumn{1}{c|}{6002}         & \multicolumn{1}{c|}{750}           & \multicolumn{1}{c|}{750}          & \multicolumn{1}{c|}{10277}           & \multicolumn{1}{c|}{34266}              & \multicolumn{1}{c|}{859}           & \multicolumn{1}{c|}{251}                  & \multicolumn{1}{c|}{235979386}                  & \multicolumn{1}{c|}{235.979386}              & \multicolumn{1}{c|}{3.279829621872166E-103}  \\ \midrule
\multicolumn{1}{|c|}{46}          & \multicolumn{1}{c|}{7500}            & \multicolumn{1}{c|}{21607}           & \multicolumn{1}{c|}{6002}         & \multicolumn{1}{c|}{750}           & \multicolumn{1}{c|}{750}          & \multicolumn{1}{c|}{10235}           & \multicolumn{1}{c|}{34064}              & \multicolumn{1}{c|}{849}           & \multicolumn{1}{c|}{31}                   & \multicolumn{1}{c|}{27638401}                   & \multicolumn{1}{c|}{27.638401}               & \multicolumn{1}{c|}{9.927826703704467E-13}   \\ \midrule
\multicolumn{1}{|c|}{47}          & \multicolumn{1}{c|}{7500}            & \multicolumn{1}{c|}{21609}           & \multicolumn{1}{c|}{6002}         & \multicolumn{1}{c|}{750}           & \multicolumn{1}{c|}{750}          & \multicolumn{1}{c|}{11377}           & \multicolumn{1}{c|}{38597}              & \multicolumn{1}{c|}{920}           & \multicolumn{1}{c|}{689}                  & \multicolumn{1}{c|}{644477962}                  & \multicolumn{1}{c|}{644.477962}              & \multicolumn{1}{c|}{1.2810988897753624E-280} \\ \midrule
\multicolumn{1}{|c|}{48}          & \multicolumn{1}{c|}{7500}            & \multicolumn{1}{c|}{21397}           & \multicolumn{1}{c|}{6002}         & \multicolumn{1}{c|}{750}           & \multicolumn{1}{c|}{750}          & \multicolumn{1}{c|}{4488}            & \multicolumn{1}{c|}{14083}              & \multicolumn{1}{c|}{815}           & \multicolumn{1}{c|}{1}                    & \multicolumn{1}{c|}{18442}                      & \multicolumn{1}{c|}{0.018442}                & \multicolumn{1}{c|}{0.981728}                \\ \midrule
\multicolumn{1}{|c|}{49}          & \multicolumn{1}{c|}{7500}            & \multicolumn{1}{c|}{21410}           & \multicolumn{1}{c|}{6002}         & \multicolumn{1}{c|}{750}           & \multicolumn{1}{c|}{750}          & \multicolumn{1}{c|}{12792}           & \multicolumn{1}{c|}{44789}              & \multicolumn{1}{c|}{1031}          & \multicolumn{1}{c|}{668}                  & \multicolumn{1}{c|}{672741572}                  & \multicolumn{1}{c|}{672.741572}              & \multicolumn{1}{c|}{6.81228495760467E-293}   \\ \midrule
\multicolumn{1}{|c|}{50}          & \multicolumn{1}{c|}{7500}            & \multicolumn{1}{c|}{21566}           & \multicolumn{1}{c|}{6002}         & \multicolumn{1}{c|}{750}           & \multicolumn{1}{c|}{750}          & \multicolumn{1}{c|}{13253}           & \multicolumn{1}{c|}{47290}              & \multicolumn{1}{c|}{851}           & \multicolumn{1}{c|}{1}                    & \multicolumn{1}{c|}{9154}                       & \multicolumn{1}{c|}{0.009154}                & \multicolumn{1}{c|}{0.990888}                \\ \midrule
\multicolumn{1}{|c|}{51}          & \multicolumn{1}{c|}{7500}            & \multicolumn{1}{c|}{20454}           & \multicolumn{1}{c|}{4502}         & \multicolumn{1}{c|}{1500}          & \multicolumn{1}{c|}{1500}         & \multicolumn{1}{c|}{11031}           & \multicolumn{1}{c|}{39763}              & \multicolumn{1}{c|}{972}           & \multicolumn{1}{c|}{1}                    & \multicolumn{1}{c|}{2151}                       & \multicolumn{1}{c|}{0.002151}                & \multicolumn{1}{c|}{0.997852}                \\ \midrule
\multicolumn{1}{|c|}{52}          & \multicolumn{1}{c|}{7500}            & \multicolumn{1}{c|}{20450}           & \multicolumn{1}{c|}{4502}         & \multicolumn{1}{c|}{1500}          & \multicolumn{1}{c|}{1500}         & \multicolumn{1}{c|}{8927}            & \multicolumn{1}{c|}{30739}              & \multicolumn{1}{c|}{855}           & \multicolumn{1}{c|}{1}                    & \multicolumn{1}{c|}{738}                        & \multicolumn{1}{c|}{7.38E-4}                 & \multicolumn{1}{c|}{0.999263}                \\ \midrule
\multicolumn{1}{|c|}{53}          & \multicolumn{1}{c|}{7500}            & \multicolumn{1}{c|}{20616}           & \multicolumn{1}{c|}{4502}         & \multicolumn{1}{c|}{1500}          & \multicolumn{1}{c|}{1500}         & \multicolumn{1}{c|}{11843}           & \multicolumn{1}{c|}{43792}              & \multicolumn{1}{c|}{894}           & \multicolumn{1}{c|}{1}                    & \multicolumn{1}{c|}{37}                         & \multicolumn{1}{c|}{3.7E-5}                  & \multicolumn{1}{c|}{0.999964}                \\ \midrule
\multicolumn{1}{|c|}{54}          & \multicolumn{1}{c|}{7500}            & \multicolumn{1}{c|}{20530}           & \multicolumn{1}{c|}{4502}         & \multicolumn{1}{c|}{1500}          & \multicolumn{1}{c|}{1500}         & \multicolumn{1}{c|}{9961}            & \multicolumn{1}{c|}{35071}              & \multicolumn{1}{c|}{1053}          & \multicolumn{1}{c|}{502}                  & \multicolumn{1}{c|}{480184105}                  & \multicolumn{1}{c|}{480.184105}              & \multicolumn{1}{c|}{2.8797108920892045E-209} \\ \midrule
\multicolumn{1}{|c|}{55}          & \multicolumn{1}{c|}{7500}            & \multicolumn{1}{c|}{20563}           & \multicolumn{1}{c|}{4502}         & \multicolumn{1}{c|}{1500}          & \multicolumn{1}{c|}{1500}         & \multicolumn{1}{c|}{9462}            & \multicolumn{1}{c|}{32930}              & \multicolumn{1}{c|}{1368}          & \multicolumn{1}{c|}{769}                  & \multicolumn{1}{c|}{739302414}                  & \multicolumn{1}{c|}{739.302414}              & \multicolumn{1}{c|}{8.45E-322}               \\ \midrule
\multicolumn{1}{|c|}{56}          & \multicolumn{1}{c|}{7500}            & \multicolumn{1}{c|}{20493}           & \multicolumn{1}{c|}{4502}         & \multicolumn{1}{c|}{1500}          & \multicolumn{1}{c|}{1500}         & \multicolumn{1}{c|}{9084}            & \multicolumn{1}{c|}{31398}              & \multicolumn{1}{c|}{833}           & \multicolumn{1}{c|}{1}                    & \multicolumn{1}{c|}{7545}                       & \multicolumn{1}{c|}{0.007545}                & \multicolumn{1}{c|}{0.992484}                \\ \midrule
\multicolumn{1}{|c|}{57}          & \multicolumn{1}{c|}{7500}            & \multicolumn{1}{c|}{20491}           & \multicolumn{1}{c|}{4502}         & \multicolumn{1}{c|}{1500}          & \multicolumn{1}{c|}{1500}         & \multicolumn{1}{c|}{4922}            & \multicolumn{1}{c|}{16088}              & \multicolumn{1}{c|}{817}           & \multicolumn{1}{c|}{1}                    & \multicolumn{1}{c|}{104472}                     & \multicolumn{1}{c|}{0.104472}                & \multicolumn{1}{c|}{0.9008}                  \\ \midrule
\multicolumn{1}{|c|}{58}          & \multicolumn{1}{c|}{7500}            & \multicolumn{1}{c|}{20594}           & \multicolumn{1}{c|}{4502}         & \multicolumn{1}{c|}{1500}          & \multicolumn{1}{c|}{1500}         & \multicolumn{1}{c|}{5943}            & \multicolumn{1}{c|}{19507}              & \multicolumn{1}{c|}{987}           & \multicolumn{1}{c|}{267}                  & \multicolumn{1}{c|}{256660486}                  & \multicolumn{1}{c|}{256.660486}              & \multicolumn{1}{c|}{3.4340775952647096E-112} \\ \midrule
\multicolumn{1}{|c|}{59}          & \multicolumn{1}{c|}{7500}            & \multicolumn{1}{c|}{20406}           & \multicolumn{1}{c|}{4502}         & \multicolumn{1}{c|}{1500}          & \multicolumn{1}{c|}{1500}         & \multicolumn{1}{c|}{9340}            & \multicolumn{1}{c|}{32356}              & \multicolumn{1}{c|}{898}           & \multicolumn{1}{c|}{158}                  & \multicolumn{1}{c|}{148111431}                  & \multicolumn{1}{c|}{148.111431}              & \multicolumn{1}{c|}{4.74472781242486E-65}    \\ \midrule
\multicolumn{1}{|c|}{60}          & \multicolumn{1}{c|}{7500}            & \multicolumn{1}{c|}{20445}           & \multicolumn{1}{c|}{4502}         & \multicolumn{1}{c|}{1500}          & \multicolumn{1}{c|}{1500}         & \multicolumn{1}{c|}{8882}            & \multicolumn{1}{c|}{30572}              & \multicolumn{1}{c|}{827}           & \multicolumn{1}{c|}{1}                    & \multicolumn{1}{c|}{14066}                      & \multicolumn{1}{c|}{0.014066}                & \multicolumn{1}{c|}{0.986033}                \\ \midrule
\multicolumn{1}{|c|}{61}          & \multicolumn{1}{c|}{10000}           & \multicolumn{1}{c|}{28613}           & \multicolumn{1}{c|}{8002}         & \multicolumn{1}{c|}{1000}          & \multicolumn{1}{c|}{1000}         & \multicolumn{1}{c|}{16234}           & \multicolumn{1}{c|}{56222}              & \multicolumn{1}{c|}{1087}          & \multicolumn{1}{c|}{1}                    & \multicolumn{1}{c|}{1904}                       & \multicolumn{1}{c|}{0.001904}                & \multicolumn{1}{c|}{0.998099}                \\ \midrule
\multicolumn{1}{|c|}{62}          & \multicolumn{1}{c|}{10000}           & \multicolumn{1}{c|}{28675}           & \multicolumn{1}{c|}{8002}         & \multicolumn{1}{c|}{1000}          & \multicolumn{1}{c|}{1000}         & \multicolumn{1}{c|}{14261}           & \multicolumn{1}{c|}{47804}              & \multicolumn{1}{c|}{914}           & \multicolumn{1}{c|}{197}                  & \multicolumn{1}{c|}{185985480}                  & \multicolumn{1}{c|}{185.98548}               & \multicolumn{1}{c|}{1.6901841317920728E-81}  \\ \midrule
\multicolumn{1}{|c|}{63}          & \multicolumn{1}{c|}{10000}           & \multicolumn{1}{c|}{28558}           & \multicolumn{1}{c|}{8002}         & \multicolumn{1}{c|}{1000}          & \multicolumn{1}{c|}{1000}         & \multicolumn{1}{c|}{13755}           & \multicolumn{1}{c|}{45717}              & \multicolumn{1}{c|}{893}           & \multicolumn{1}{c|}{1}                    & \multicolumn{1}{c|}{43}                         & \multicolumn{1}{c|}{4.3E-5}                  & \multicolumn{1}{c|}{0.999957}                \\ \midrule
\multicolumn{1}{|c|}{64}          & \multicolumn{1}{c|}{10000}           & \multicolumn{1}{c|}{28738}           & \multicolumn{1}{c|}{8002}         & \multicolumn{1}{c|}{1000}          & \multicolumn{1}{c|}{1000}         & \multicolumn{1}{c|}{13370}           & \multicolumn{1}{c|}{44343}              & \multicolumn{1}{c|}{882}           & \multicolumn{1}{c|}{1}                    & \multicolumn{1}{c|}{127}                        & \multicolumn{1}{c|}{1.27E-4}                 & \multicolumn{1}{c|}{0.999874}                \\ \midrule
\multicolumn{1}{|c|}{65}          & \multicolumn{1}{c|}{10000}           & \multicolumn{1}{c|}{28752}           & \multicolumn{1}{c|}{8002}         & \multicolumn{1}{c|}{1000}          & \multicolumn{1}{c|}{1000}         & \multicolumn{1}{c|}{15537}           & \multicolumn{1}{c|}{53105}              & \multicolumn{1}{c|}{917}           & \multicolumn{1}{c|}{643}                  & \multicolumn{1}{c|}{606121928}                  & \multicolumn{1}{c|}{606.121928}              & \multicolumn{1}{c|}{5.826520007473361E-264}  \\ \midrule
\multicolumn{1}{|c|}{66}          & \multicolumn{1}{c|}{10000}           & \multicolumn{1}{c|}{28803}           & \multicolumn{1}{c|}{8002}         & \multicolumn{1}{c|}{1000}          & \multicolumn{1}{c|}{1000}         & \multicolumn{1}{c|}{9981}            & \multicolumn{1}{c|}{32065}              & \multicolumn{1}{c|}{852}           & \multicolumn{1}{c|}{1}                    & \multicolumn{1}{c|}{796}                        & \multicolumn{1}{c|}{7.96E-4}                 & \multicolumn{1}{c|}{0.999205}                \\ \midrule
\multicolumn{1}{|c|}{67}          & \multicolumn{1}{c|}{10000}           & \multicolumn{1}{c|}{28632}           & \multicolumn{1}{c|}{8002}         & \multicolumn{1}{c|}{1000}          & \multicolumn{1}{c|}{1000}         & \multicolumn{1}{c|}{13418}           & \multicolumn{1}{c|}{44550}              & \multicolumn{1}{c|}{861}           & \multicolumn{1}{c|}{448}                  & \multicolumn{1}{c|}{439405919}                  & \multicolumn{1}{c|}{439.405919}              & \multicolumn{1}{c|}{1.4772121624185204E-191} \\ \midrule
\multicolumn{1}{|c|}{68}          & \multicolumn{1}{c|}{10000}           & \multicolumn{1}{c|}{28830}           & \multicolumn{1}{c|}{8002}         & \multicolumn{1}{c|}{1000}          & \multicolumn{1}{c|}{1000}         & \multicolumn{1}{c|}{17774}           & \multicolumn{1}{c|}{63650}              & \multicolumn{1}{c|}{874}           & \multicolumn{1}{c|}{1}                    & \multicolumn{1}{c|}{3047}                       & \multicolumn{1}{c|}{0.003047}                & \multicolumn{1}{c|}{0.996959}                \\ \midrule
\multicolumn{1}{|c|}{69}          & \multicolumn{1}{c|}{10000}           & \multicolumn{1}{c|}{28717}           & \multicolumn{1}{c|}{8002}         & \multicolumn{1}{c|}{1000}          & \multicolumn{1}{c|}{1000}         & \multicolumn{1}{c|}{14505}           & \multicolumn{1}{c|}{48831}              & \multicolumn{1}{c|}{861}           & \multicolumn{1}{c|}{1}                    & \multicolumn{1}{c|}{1691}                       & \multicolumn{1}{c|}{0.001691}                & \multicolumn{1}{c|}{0.998311}                \\ \midrule
\multicolumn{1}{|c|}{70}          & \multicolumn{1}{c|}{10000}           & \multicolumn{1}{c|}{28604}           & \multicolumn{1}{c|}{8002}         & \multicolumn{1}{c|}{1000}          & \multicolumn{1}{c|}{1000}         & \multicolumn{1}{c|}{16032}           & \multicolumn{1}{c|}{55089}              & \multicolumn{1}{c|}{855}           & \multicolumn{1}{c|}{1}                    & \multicolumn{1}{c|}{436}                        & \multicolumn{1}{c|}{4.36E-4}                 & \multicolumn{1}{c|}{0.999564}                \\ \midrule
\multicolumn{1}{|c|}{71}          & \multicolumn{1}{c|}{10000}           & \multicolumn{1}{c|}{27114}           & \multicolumn{1}{c|}{6002}         & \multicolumn{1}{c|}{2000}          & \multicolumn{1}{c|}{2000}         & \multicolumn{1}{c|}{15244}           & \multicolumn{1}{c|}{55476}              & \multicolumn{1}{c|}{2286}          & \multicolumn{1}{c|}{652}                  & \multicolumn{1}{c|}{652324945}                  & \multicolumn{1}{c|}{652.324945}              & \multicolumn{1}{c|}{5.016628484164324E-284}  \\ \midrule
\multicolumn{1}{|c|}{72}          & \multicolumn{1}{c|}{10000}           & \multicolumn{1}{c|}{27515}           & \multicolumn{1}{c|}{6002}         & \multicolumn{1}{c|}{2000}          & \multicolumn{1}{c|}{2000}         & \multicolumn{1}{c|}{10588}           & \multicolumn{1}{c|}{36029}              & \multicolumn{1}{c|}{867}           & \multicolumn{1}{c|}{1}                    & \multicolumn{1}{c|}{15974}                      & \multicolumn{1}{c|}{0.015974}                & \multicolumn{1}{c|}{0.984154}                \\ \midrule
\multicolumn{1}{|c|}{73}          & \multicolumn{1}{c|}{10000}           & \multicolumn{1}{c|}{27411}           & \multicolumn{1}{c|}{6002}         & \multicolumn{1}{c|}{2000}          & \multicolumn{1}{c|}{2000}         & \multicolumn{1}{c|}{9596}            & \multicolumn{1}{c|}{32332}              & \multicolumn{1}{c|}{862}           & \multicolumn{1}{c|}{422}                  & \multicolumn{1}{c|}{440653751}                  & \multicolumn{1}{c|}{440.653751}              & \multicolumn{1}{c|}{4.240514855635819E-192}  \\ \midrule
\multicolumn{1}{|c|}{74}          & \multicolumn{1}{c|}{10000}           & \multicolumn{1}{c|}{27271}           & \multicolumn{1}{c|}{6002}         & \multicolumn{1}{c|}{2000}          & \multicolumn{1}{c|}{2000}         & \multicolumn{1}{c|}{15985}           & \multicolumn{1}{c|}{59167}              & \multicolumn{1}{c|}{873}           & \multicolumn{1}{c|}{1}                    & \multicolumn{1}{c|}{2033}                       & \multicolumn{1}{c|}{0.002033}                & \multicolumn{1}{c|}{0.997969}                \\ \midrule
\multicolumn{1}{|c|}{75}          & \multicolumn{1}{c|}{10000}           & \multicolumn{1}{c|}{27228}           & \multicolumn{1}{c|}{6002}         & \multicolumn{1}{c|}{2000}          & \multicolumn{1}{c|}{2000}         & \multicolumn{1}{c|}{13506}           & \multicolumn{1}{c|}{47651}              & \multicolumn{1}{c|}{2223}          & \multicolumn{1}{c|}{621}                  & \multicolumn{1}{c|}{639112478}                  & \multicolumn{1}{c|}{639.112478}              & \multicolumn{1}{c|}{2.7423451190246526E-278} \\ \midrule
\multicolumn{1}{|c|}{76}          & \multicolumn{1}{c|}{10000}           & \multicolumn{1}{c|}{27345}           & \multicolumn{1}{c|}{6002}         & \multicolumn{1}{c|}{2000}          & \multicolumn{1}{c|}{2000}         & \multicolumn{1}{c|}{12066}           & \multicolumn{1}{c|}{41598}              & \multicolumn{1}{c|}{1253}          & \multicolumn{1}{c|}{326}                  & \multicolumn{1}{c|}{307525901}                  & \multicolumn{1}{c|}{307.525901}              & \multicolumn{1}{c|}{2.779537506735469E-134}  \\ \midrule
\multicolumn{1}{|c|}{77}          & \multicolumn{1}{c|}{10000}           & \multicolumn{1}{c|}{27310}           & \multicolumn{1}{c|}{6002}         & \multicolumn{1}{c|}{2000}          & \multicolumn{1}{c|}{2000}         & \multicolumn{1}{c|}{10310}           & \multicolumn{1}{c|}{34812}              & \multicolumn{1}{c|}{835}           & \multicolumn{1}{c|}{1}                    & \multicolumn{1}{c|}{10970}                      & \multicolumn{1}{c|}{0.01097}                 & \multicolumn{1}{c|}{0.989091}                \\ \midrule
\multicolumn{1}{|c|}{78}          & \multicolumn{1}{c|}{10000}           & \multicolumn{1}{c|}{27306}           & \multicolumn{1}{c|}{6002}         & \multicolumn{1}{c|}{2000}          & \multicolumn{1}{c|}{2000}         & \multicolumn{1}{c|}{12092}           & \multicolumn{1}{c|}{41711}              & \multicolumn{1}{c|}{1004}          & \multicolumn{1}{c|}{228}                  & \multicolumn{1}{c|}{218680041}                  & \multicolumn{1}{c|}{218.680041}              & \multicolumn{1}{c|}{1.0684631282749114E-95}  \\ \midrule
\multicolumn{1}{|c|}{79}          & \multicolumn{1}{c|}{10000}           & \multicolumn{1}{c|}{27315}           & \multicolumn{1}{c|}{6002}         & \multicolumn{1}{c|}{2000}          & \multicolumn{1}{c|}{2000}         & \multicolumn{1}{c|}{14069}           & \multicolumn{1}{c|}{50130}              & \multicolumn{1}{c|}{848}           & \multicolumn{1}{c|}{1}                    & \multicolumn{1}{c|}{1447}                       & \multicolumn{1}{c|}{0.001447}                & \multicolumn{1}{c|}{0.998555}                \\ \midrule
\multicolumn{1}{|c|}{80} & \multicolumn{1}{c|}{10000} & \multicolumn{1}{c|}{27375} & \multicolumn{1}{c|}{6002} & \multicolumn{1}{c|}{2000} & \multicolumn{1}{c|}{2000} & \multicolumn{1}{c|}{14851} & \multicolumn{1}{c|}{53699} & \multicolumn{1}{c|}{859}  & \multicolumn{1}{c|}{1}    & \multicolumn{1}{c|}{180}       & \multicolumn{1}{c|}{1.8E-4}     & \multicolumn{1}{c|}{0.999821}                \\ \bottomrule
\end{tabular}
 \vspace{0.5cm}
\caption{Benchmark description - Cases 41 to 80}
\label{tab:part2}
\end{table*}

\bibliographystyle{myIEEEtran}
\bibliography{IEEEabrv,doc.bib}

\end{document}